\documentclass[12pt]{article}
\usepackage{amssymb,amsmath}
\begin{document}

\begin{titlepage}

                            \begin{center}
                            \vspace*{2cm}
\large\bf{Weak Chaos from Tsallis Entropy}\\

                            \vfill

              \normalsize\sf    NIKOS \  KALOGEROPOULOS $^\ast$\\

                            \vspace{0.2cm}
                            
 \normalsize\sf Weill Cornell Medical College in Qatar\\
 Education City,  P.O.  Box 24144\\
 Doha, Qatar\\

                            \end{center}

                            \vfill

                     \centerline{\normalsize\bf Abstract}
                     
                           \vspace{3mm}
                     
\normalsize\rm\setlength{\baselineskip}{18pt} 

\noindent We present a geometric, model-independent, argument that aims to explain why the Tsallis 
entropy describes systems exhibiting ``weak chaos", namely systems whose underlying dynamics has 
vanishing largest Lyapunov exponent. Our argument relies on properties of a deformation map of the reals
induced by the Tsallis entropy, and its conclusion agrees with all currently known results. \\    

                             \vfill

                             
                             \vfill

\noindent\rule{8cm}{0.2mm}\\
   \noindent \small\rm $^\ast$  E-mail: \ \  \small\rm nik2011@qatar-med.cornell.edu\\

\end{titlepage}


                            \newpage

\normalsize\rm\setlength{\baselineskip}{18pt}

          \centerline{\large\sc 1. \ Introduction}

                                     \vspace{5mm}

The Tsallis entropy is a functional first introduced in the Physics literature in 1988 [1], providing 
an alternative to the Boltzmann/Gibbs/Shannon (BGS) entropy. It aims to describe collective phenomena
with  long-range spatial and temporal correlations, systems whose phase portraits exhibit a fractal-like behavior 
for which there is no reason why the description through the BGS entropy should be accurate or even valid [2] (and references therein).   
Although the dynamical basis of the Tsallis entropy remains unclear so far, not totally unlike that of the BGS entropy [2]-[6],
there is certain progress that has been made in identifying characteristics of systems that seem to be well described by it. 
One class of such systems are ones exhibiting ``weak chaos" or being at the ``edge of chaos" for some subset of 
their parameter space, more concretely referring to systems with vanishing largest Lyapunov exponent [2], [7]-[10].\\

A general and relatively formal justification of why such systems are described by the Tsallis entropy 
has been lacking so far [2],[11]. This is  the issue that we are addressing in the present work.
To be more specific, we present an argument about the converse: if a system is described 
by the Tsallis entropy then its largest  Lyapunov exponent vanishes. This fact is in agreement with all currently known 
results pertaining to the microscopic dynamics of systems described by the Tsallis entropy. As a by-product we also comment on modified 
definitions of asymptotic numerical invariants of dynamical systems better reflecting the properties of the Tsallis entropy [2], [11]-[16].  
Our approach relies on the metric aspects of Riemannian manifolds..  
For this reason some arguments can also be applied, or non-trivially generalized, in the much more  general classes of (geodesic) metric spaces, 
such as spaces with a (negative) upper bound \  $k<0$ \ on their curvature,  known as \ $CAT(k)$ \ spaces.  
A feature of our presentation that sets it apart from most other descriptions of Riemannian methods addressing a Physics audience, is 
that it stresses, as much as possible the synthetic, rather than the analytical viewpoint. The advantage of this approach is that it is
immediately generalizable to more general geometric structures such as the \ $CAT(k)$ \ spaces, mentioned in this work, that do not have
very ``nice" smoothness/regularity properties.     
In Section 2, we provide a very brief background in Riemannian geometry in an attempt to make the exposition at least minimally 
self-contained for our audience. We forego essentially all details and infinitesimal calculations as they can be found in any exposition 
of Riemannian geometry, such as [17],[18]. In Section 3 we present our argument. Section 4 points toward possible extensions of the geometric 
structures and future research.\\

                                  \vspace{5mm}


          \centerline{\large\sc 2. \  Geometric preliminaries}

                                        \vspace{3mm}

Riemannian manifolds \ $M$ \ of dimension \ $n$ \ are structures that are made up of local patches of Euclidean spaces of 
(integer Hausdorff) dimension \ $n$. \ Each of these patches can be parametrized by a coordinate system. The patches are ``glued"
together via functions having desired regularity properties (homeomorphisms, PL-equivalences, diffeomorphisms etc). 
For most physical applications Riemannian manifolds are assumed to be smooth enough, namely the transition functions that glue together 
local patches of \ $\mathbb{R}^n$ \ are assumed to be \ $C^\infty$. \ Riemannian manifolds are therefore metric spaces whose metric is locally Euclidean
since it is descended from the Euclidean metric of each patch. Quite importantly, they 
are length spaces and even more specifically, they are geodesic spaces (also called intrinsic length spaces), namely spaces for which the distance 
between any two of their points is realized as the infimum of  the lengths of all piecewise smooth curves 
joining them. A direct consequence  is that Riemannian manifolds have built-in the concept of direction, as can be seen by the fact that their metric 
structure is determined by specifying an ellipsoid in the neighborhood of each of their points. This ellipsoiid is a reminder of the Euclidean patches
that were put together to form the metric  of Riemannian manifolds. An ellipsoid is analytically described as the unit sphere of a positive-definite, 
non-degenerate quadratic form \ $g$, \ called the metric tensor, which is therefore defined though its sections with each 2-dimensional subspace of the 
tangent space \ $T_xM$ \ at each point \ $x \in M$. \ The tangent space \ $T_xM$ \ is the set of vectors  that are tangent at \ $x\in M$ \ to all curves within \  $M
$ \ passing through \ $x$. \ Hence, roughly 
speaking, elements of \ $T_xM$ \ are derivations in the space of smooth real functions defined on the neighborhood of each \ $x\in M$. \ Equivalently,  the 
curves passing through \ $x$ \ are the integral curves of the  corresponding elements of \ $T_x M$. \ One can map each element \ $X\in T_xM$ \ to its 
corresponding integral curve \ $c(t)$ \ by using the exponential map \ $\exp_x : T_xM \rightarrow M$ \  defined by \ $\exp_x(X)(t) = c(t)$. \\
  
The fact that the two metric structures, one which is induced by the underlying topology of \ $\mathbb{R}^n$ \ and the other which is induced by the length 
structure are equivalent, is non-trivial, and is the content of the Hopf-Rinow theorem [17],[18]. In that theorem it is assumed that the Riemannian 
manifold is complete, meaning that  no points have been eliminated from it, as for instance in the case of \ $\mathbb{R}^2 \backslash \{0 \}$ \ 
where \ $ 0 $ \ denotes the origin. This is more compactly expressed by requiring \ $\exp_x$ \ to be complete. The process of actually requiring the 
exponential map to be complete may appear quite technical and far removed from the needs of Physics, but such a statement would only be partly correct. 
There is at a least one notable case of  use of non-complete 
manifolds for physical purposes and this the case of the singularity determination  in black holes. 
In the present work we will ignore cases of non-complete manifolds since they may be encountered rarely in Statistical Mechanics 
so they are not very important for our purposes.\\
 
  It may also be noticed that there is nothing really ``sacred" about using 
 a non-degenerate quadratic form, namely an ellipsoid, to determine the distance function as is done in Euclidean/Riemannian spaces. 
 One could use an appropriately scaled version of  any symmetric convex body instead, which is the unit ellipsoid of a general Banach norm. 
 This is indeed done in the case of Finsler manifolds. 
 The general determination  of a symmetric convex body, however, requires infinite parameters, even in the planar  case. By contrast, the determination of 
 an ellipsoid requires knowledge of just the lengths of its axes. Moreover, no other Banach spaces except the 
 square summable/integrable cases are induced from an inner product. The inner product  is a very well-understood and convenient technical device, 
 so we would like to keep using it in our models, unless particular requirements force us to abandon it.\\

As in any metric spaces, also in the case of Riemannian manifolds \ $M$, \ one would like to find ways of determining the distances between any 
two of their points. This turns out to be a hopeless task, even in principle. First of all, to appropriately determine a distance \ $d(x,y), \ \ x,y \in M$, \ the 
corresponding Riemannian metric \ $g$ \ will have to be determined. This amounts to determining \ $n(n+1)/2$ \ numbers at every point of an $n$-
dimensional Riemannian manifold as \ $g$ \ is a symmetric, non-degenerate quadratic form. This is a local, even though quite often cumbersome, calculation 
so it is feasible. Determining, however,  the minimum distance 
over all (rectifiable) curves between any two points \ $x$ \ and \ $y$ \  is an intractable task. We can try, for instance, to find the distance between two matrices
belonging to the (linear)  representation of a Lie group, whose set is a Riemannian manifold when endowed with the Cartan-Killing metric. 
This turns out to be not a very simple task.  
To see how things can get even harder, consider  the following example: let  \ $N$ \ be a Riemanian manifold which is 
not simply connected, meaning that there are loops in \ $N$ \ that are not continuously deformable (``homotopic") to points.
It turns out that each such \ $N$ \  has a ``universal cover" \ $\tilde{N}$ \ which is simply-connected. Although globally \ $N$ \ and \ $\tilde{N}$ \  may be 
appear different, they are locally (arc-wise) isometric, in the neighborhood of (almost) every point. Isometry signifies that the distances are locally preserved 
in going from \ $N$ \ to \ $\tilde{N}$. \  To be more specific, the length structure lifts from \ $N$ \ to \ $\tilde{N}$ \ by assigning 
to each curve of \ $\tilde{N}$ \ its image in \ $N$. \ What is still unclear is how to determine the metric on \ $\tilde{N}$. \ 
Someone may ask, for instance,  whether the diameter of \ $\tilde{N}$ \ defined as \ $\sup d(x,y), \ \ \forall \  x, y \in \tilde{N}$ \ is finite or not. This reduces to 
deciding whether the fundamental group \ $\pi_1(N)$, \ which is an algebraic structure (a group) measuring the algebraic complexity of families of homotopic 
loops in \ $N$, \  is finite or not. This is however an undecidable proposition, therefore the diameter of \ $\tilde{N}$ \ cannot be effectively calculated [19],[20].\\
                   
To get around these difficulties, one would like to define effectively computable invariants of \ $M$ \ which are local in the sense that they are defined for a 
neighborhood of each point of \ $M$. \  Such invariants are usually related to the Riemann curvature tensor, which is a relatively complicated non-linear 
expression of the metric tensor \ $g$ \ its inverse \  $g^{-1}$ \ and their second derivatives [18],[19].  
As such an expression is not very illuminating and it manifestly 
depends on the system of coordinates employed, one may wish to specify it in an alternative way, which would exhibit its linear-algebraic character and 
which would make more prominent its coordinate independence, and consequently its geometric meaning. This is a partial motivation for introducing the 
concept of a Levi-Civita connection [17],[18]. A connection, as the word indicates, is a way of splitting the tangent space of a Riemannian
manifold into a horizontal and a vertical part \ $T_xM = (T_xM)^\top + (T_xM)^\bot $ \  at each point \ $x\in M$. \  This is an orthogonal splitting 
which is required to be compatible with the properties of \ $g$. \ The horizontal part of this splitting is used to compare or ``connect" the tangent spaces 
over different points of \ $M$, \ hence the name of the structure. 
The connection amounts to essentially choosing how to identify the tangent spaces \ $T_xM$ \ at different points \ $x\in M$ \ by declaring which curves are 
horizontal, by locally specifying their tangent vectors. Hence, a connection can be also seen as a derivation on \ $T_xM$, \ or equivalently as the directional 
(covariant) 
derivative of vector fields \ $Y$ \ along curves with tangents determined by the vector field \ $X$. \ The notation for this is \ $\nabla_XY$. \  
This (covariant) derivative operation  can be extended to  tensor products of  \ $T_xM$ \ and its duals \ $T_x^\ast M$ \ by linearity and Leibniz's rule. 
The Riemann curvature tensor expresses the path dependence (or lack of integrability) of the transport of a vector around an infinitesimal loop. 
To determine it, one transports a  vector \ $Z \in 
T_xM $ \  around the loop with the shape of an  infinitesimal rectangle whose sides are the vector fields \ $X, Y \in TM$. \ 
Following this motivation and wanting to  keep the geometric character (reparametrization independence) of the resulting object by requiring it to be a tensor, the Riemann tensor is defined as  
\begin{equation}       
 R(X,Y)Z = \nabla_X\nabla_Y Z - \nabla_Y\nabla_X Z - \nabla_{[X,Y]} Z
\end{equation}
Here, the square brackets indicate a commutator. 
From this relation one immediately sees that the Riemann tensor is the obstruction to the integrability of the vector \ $Z$ \  across the distribution determined 
by \ $X$ \ and \ $Y$  \ and, as such, it quantifies curvature. Hence, the curvature is  fundamentally a 2-dimensional 
concept, although it appears far more complicated when seen in the totality of all 2-dimensional subspaces of \ $T_xM$, \ formally known as the Grassmann
manifold \ $G_{n,2}(M)$.  \ To quantify more concretely and geometrically the concept of curvature, several scalar quantities derived from 
the Riemann tensor are utilized [17]-[20]. The most general of the widely used ones, is the sectional curvature \ $K$, \ which is defined in each 2-dimensional 
subspace of $T_xM$. \ Let such a subspace \ $\Sigma$ \ be spanned by the orthonormal vectors \ $e_1, e_2$ \ with respect to the Riemannian metric \ $g$. \ 
Then
\begin{equation}
      K_\Sigma = g(R(e_1, e_2)e_2, e_1)
\end{equation}
The geometric interpretation of \ $K_\Sigma$ \ is provided by the following well-known fact [18],[20]. 
Consider a circle \ $c_r$ \ of radius \ $r$ \ in \ $\Sigma$. \ Use the 
exponential map \ $\exp$ \ to map this circle to a closed curve of length \ $l_r$ \ in \ $M$. \   Then 
\begin{equation}
    K_\Sigma = \frac{3}{\pi} \lim_{r\rightarrow 0} \frac{2\pi r - l_r}{r^3}
\end{equation}
We immediately see that positive sectional curvature corresponds to the actual shrinking of the length \ $l_r$ \ of \ $c_r$, \ whereas negative sectional 
curvature is equivalent to an expansion of \ $l_r$. \ This is the critical geometric observation that will be used in the sequel and on which our 
whole argument relies. Conversely, if one knows the sectional curvature along all 2-dimensional planes at a point  \ $x\in M$, \ then one can determine the 
Riemann tensor by the algebraic operation of polarization [17],[18]. 
It may be worth mentioning at this point that the definitions of the Riemann tensor through the 
connection or through a metric are equivalent. This happens because Riemannian manifolds have the following nice property: there is a unique symmetric
connection, which is compatible with the metric \ $g$, \ meaning that it preserves the inner products of any two vectors \ $X,Y \in T_xM$. \  As a reminder,
a connection is symmetric, also called torsion-free, if it satisfies  \ $\nabla_XY - \nabla_YX = [X, Y]$, \ so in a coordinate basis the corresponding connection 
coefficients (Christoffel symbols) are symmetric.  For completeness, 
we state the explicit expression of the connection in terms of the metric $g$, whose construction was inspired by group cohomology, 
and is due to Koszul
\begin{equation}   
  2 g(\nabla_XY, Z) =  X[g(Y,Z)] - Z[g(X,Y)] + Y[g(X,Z)] - g(X,[Y,Z]) + g(Z,[X,Y]) + g(Y,[Z,X])
\end{equation}
In Riemannian geometry a curve is called geodesic between two points \ $x,y\in M$, \ if it is a stationary point of the length functional on all (rectifiable)
curves joining \ $x$ \ and \ $y$. \ The concept of geodesic reflects the Euclidean concept of the ``most straight line" between the two points. This line does not 
have to be a shortest one though, or even to be unique. Consider, for instance, a major circle joining two points of a 2-sphere. It is a geodesic even though 
it may not be minimal. Moreover, if these two points are the south and the north poles then this geodesic is only one member of an infinite continuous family.  
From its variational characterization, or as a definition, it turns out that a geodesic whose tangent is \ $X\in TM$ \ can be analytically determined as the 
solution of 
\begin{equation}
  \nabla_XX = 0 
\end{equation}
which states that the tangent vector \ $X$ \ to the curve remains parallel to itself, hence the corresponding integral curve is the ``straightest possible" line 
joining \ $x$ \  and \ $y$ \ [17],[18]. In  more physical terms, the equation defining geodesics is the 
Euler-Lagrange equation of the kinetic energy of a free point particle moving on \ $M$. \ Therefore the geodesics of \ $M$ \ are the trajectories of freely 
moving particles on \ $M$. \ We have tacitly assumed in the preceding comments that all geodesics are parametrized by arc-length. \\  

Another fundamental concept in Riemannian geometry is that of  Jacobi fields. The idea is that these are vector fields joining nearby geodesics. As such, 
Jacobi fields describe how nearby geodesics converge or diverge with respect to each other. They are the means by which gravitational interactions in
General Relativity are determined (albeit on manifolds with a ``metric" of indefinite/Lorentzian signature). They quantify the tidal forces exerted on an 
object by gravity. So, the Jacobi fields can be variationally characterized as extremizing the equation of geodesics. 
In more physical terms, this amounts to setting the second derivative of the kinetic energy of a free particle to zero. 
To find such an extremum one differentiates the geodesic equation using variations of via a family of interpolating geodesics. 
Then the equation for Jacobi fields is derived, known as the Jacobi or geodesic deviation equation. It is
\begin{equation}
   \nabla_X\nabla_X J + R(J, X)X = 0
\end{equation}
where \ $X$ \ is tangent to the geodesic, and where \ $J$ \ indicates a Jacobi field [17].[18]. 
We see that the Jacobi equation is a  linearization of the equation for geodesics. 
It is extremely difficult to explicitly solve, in practice, either the 
geodesic or the Jacobi equation, except in few cases of particularly simple Riemannian manifolds. Such a class of simple manifolds, which are used as 
prototypes/models for comparison with most other cases, are provided by the manifolds of constant sectional curvature. 
To keep the topological complications to a minimum, it is the simply connected ones (universal covers of all the others) that are used as such models. 
They are called space forms. The idea of comparison geometry is the 
same as in physics: one tries to understand some simple model cases and then uses the conclusions reached from them to compare the results 
with more general spaces or realistic systems. Space forms are simple enough that one can actually explicitly calculate their sectional curvatures \
 $K$ \ [17],[18] and find 
\begin{equation}
     R(X,Y)Z = K (g(Y,Z)X - g(Z,X)Y) 
\end{equation}
Then following the Cartan-Ambrose-Hicks theorem one can prove that two such space forms  having equal sectional curvatures are isometric [17],[18].
This is a very concrete example that shows that curvature indeed determines the metric properties of a manifold which, after all, was a motivation 
behind the introduction of the Riemann curvature tensor [19],[20].  We see that for a manifold with constant negative sectional curvature \ 
$K<0$, \  the Jacobi equation has the general solution       
\begin{equation}
    J(t) = \sum_i^n \left\{ a_i \cosh (\sqrt{-K} \ t) + b_i \sinh (\sqrt{-K} \ t) \right\} \ e_i(t) 
\end{equation}
where \ $a_i, \ b_i$ \ are constants, \ $\{ e_i(t) \} $ \ are parallel orthonormal vectors (Fermi basis) \ and \ $t$ \ is the parameter of the geodesic 
whose  tangent is \ $X(t)$. \ We observe therefore that for a manifold with constant negative curvature, nearby geodesics separate from 
each other exponentially, asymptotically.  By comparison, if the manifold is flat namely if the Riemann tensor/sectional curvature vanishes identically, 
then the separation between nearby geodesics is linear as can be seen again from the Jacobi equation. It is also remarkable that there is nothing in 
between these behaviors, for large values of \ $t$. \ Geodesics of Riemannian manifolds either separate linearly ($K=0$) or exponentially ($K<0$). 
No other asymptotic rate of separation of nearby geodesics is possible, such as a polynomial (power-law) separation with exponent different from one, 
a fact that is  obvious from (8).\\  

                                                          \vspace{5mm}


\centerline{\large\sc 3. \  Zero Lyapunov exponents and Tsallis entropy  }

                                                            \vspace{3mm}

Ideally, one would like to be able to derive the Boltzmann/Gibbs/Shannon  form of entropy from the dynamics of the 
underlying system. This is fundamental in Boltzmann's approach to systems with many degrees of freedom [3]-[6], although it should be noted that such 
viewpoint is not necessarily shared by Gibbs [21]-[23]. It is clear that this goal of Boltzmann has not been realized yet [2],[5],[6]. The process, however, has 
given a considerable impetus to the development of measure-theoretical aspects of dynamical systems such as ergodic theory [24],[25]. 
An important realization is that the description of systems whose thermodynamic behavior is described the BGS entropy depends on whether they possess 
any (evolution) invariant measures, known as ensembles, in phase space. Such measures, whose building blocks are called ergodic measures, 
provide the stationary probability distributions of equilibrium statistical mechanics. The basic idea, due to Gibbs, is that the classical ensembles
give equivalent thermodynamic descriptions [21]. This however turns out not to be true [26]-[28] generally, 
and especially if the systems possess long-range interactions [26],[27].   
A broad set of examples of systems possessing ergodic measures are provided by the uniformly hyperbolic (Anosov) systems [24],[25]. 
Uniformly hyperbolic dynamical systems are ones for which there is a decomposition of the tangent space at each point, varying continuously but not 
necessarily smoothly, to two transversal distributions of exponentially distance-expanding and distance-contracting subspaces of the tangent space.
To be more concrete, a flow [24], [25], is a one-parameter family of diffeomorphisms of the Riemannian manifold (``phase space") \ $(M, g)$ \ 
indicated by  \ $f_t: M\rightarrow M$. \  Let the vector field generating this flow be indicated by  
\begin{equation}
   X(t) = \frac{d}{dt}(f_t(x))|_{t=0}
\end{equation}
This flow is called Anosov, if \ $\forall \ x\in M$ \ there exists a direct-sum decomposition \ $T_xM = E^s(x) \oplus E^0(x) \oplus E^u(x)$ 
\ and constants \ $c>0$ \ and \
$0<\lambda<1<\mu$ \ such that \ $\forall x\in M$ \ the following four properties hold
\begin{enumerate} 
\item $E^0(x)$ \  is a one-dimensional subspace generated by \ $X(t)$
\item $E^s(x)$ \ and \ $E^u(x)$ \ remain invariant under the flow, namely \ $d_x(f_tE^s(x)) = E^s(f_t(x))$ \ and \ $d_x(f_tE^u(x)) = E^u(f_t(x))$  
\item All \ $Y \in E^s(x)$ \ are uniformly contracting, namely \ $||d_x(f_tY)|| \leq c\lambda^t ||Y||$, \ for \ $t\geq 0$
\item All \ $Z \in E^u(x)$ \ are uniformly expanding, namely \ $||d_x(f_{-t}Z)|| \leq c\mu^{-t}||Z||$, \ for \ $t\geq 0$   
\end{enumerate}
To determine the exponential rate of expansion and contraction of distances on \ $E^u(x)$ \ and \ $E^s(x)$ \ aptly named unstable and stable spaces, 
of an Anosov system, several quantities have been introduced. Probably the simplest of them is the set of the system's Lyapunov exponents. 
This set expresses the asymptotic instability of the evolution of the system under infinitesimal perturbations. 
In other words, the Lyapunov exponents quantify how fast the infinitesimal perturbations grow or contract in time.
Let  \ $Y \in TM$ \ and let its norm with respect to the Riemannian metric \ $g$ \ be  \ $ ||Y|| = \sqrt{g(Y,Y)} $. \ 
The Lyapunov exponent of  the perturbation in the direction of \ $Y$ \ at \  $x\in M$ \  along the evolution (trajectory) 
of the system determined by \ $f_t$ \ is defined by
 \begin{equation}    
      \chi_x(Y) = \lim_{t\rightarrow\infty} \frac{1}{t} \log ||d_x(f_tY)||
\end{equation}
From their definition, it turns out that Anosov flows having at least one positive Lyapunov exponent, exhibit exponential sensitivity to the initial conditions, 
which is one definition of (strong) chaotic behavior [2],[24],[25].  When this behavior is combined with the recurrence properties on their (assumed, for 
simplicity to be compact) phase space, it results in ergodic behavior which ultimately provides the known, though not yet completely convincing, dynamical 
reasons for use of the BGS entropy in  describing the system's stationary, or nearly stationary, behavior [24],[25]. The way the BGS entropy appears in this 
context is as the \ $t\rightarrow\infty$ \ limit of  another characteristic of the system known as its Kolmogorov-Sinai (KS) entropy [24],[25]. The KS entropy is 
defined in a way similar to that of the BGS entropy, after which it is modelled, where in addition the information theoretic ancestry of it also becomes 
apparent  [24],[25].  The field of hyperbolic dynamics is replete with important technicalities, as such flows generically result in non-smoothly varying 
continuous unstable and stable distributions, which are the sub-bundles of TM generated from \ $E^u(x)$ \ and \  $E^s(x)$ \ through \ $f_t$. \ 
We will however ignore  such issues in the sequel and we will pretend instead, that the definitions that we use have the required degree of smoothness and 
refer to standard  references such as [24],[25] for most of the non-trivial regularity aspects. Various generalizations of Anosov flows exist, for instance when 
the norm bounds are not uniform or when the center space, generalizing $E^0(x)$, is multidimensional and when within which sub-exponential expansions 
and contractions are allowed [24],[25].  However the fundamental ideas behind the definition of Anosov flows stated above remain essentially unchanged.  \\  

The most common, arguably, set of examples of such uniformly hyperbolic (Anosov) systems is the geodesic flow in the tangent bundle of a manifold 
of negative sectional curvature [29]. To see that this Anosov flow indeed possesses at least one positive Lyapunov exponent is easiest in the case
that  \  $(M, g)$ \ has constant negative curvature. Then the desired result follows straightforwardly from (8) when \ $t\rightarrow\infty$: \ 
since nearby geodesics deviate from each other exponentially in the direction of a vector \ $Y$, \  the corresponding Lyapunov exponent which quantifies this 
exponential deviation, according to (8) must be positive. Proving this for the geodesic flow on a manifold of non-constant negative sectional curvature 
is non-trivial though. Such a  statement can be made plausible by employing the Rauch comparison theorem [17]: it, roughly, states that for two Riemannian 
manifolds \ $M_1$ \ and \ 
$M_2$ \ of not necessarily the same dimension, if the sectional curvatures along all corresponding 2-dimensional planes \ $\Sigma_1$ \ and \  $\Sigma_2$ \ 
at each  corresponding set of points obey \  $K_{\Sigma_1} \geq K_{\Sigma_2}$, \ then the corresponding Jacobi fields \ $J_1$ \ and \ $J_2$ \ which are 
normal to the respective radial  geodesics, which are assumed to be arc-length parametrized, obey \ $ ||J_2(t)|| \geq ||J_1(t)||. $   \  This essentially means that 
when the sectional curvature has a positive value then the radial geodesics tend to strongly converge/focus, a fact which is expected/natural if one thinks  
in terms of  General Relativity or any other generally covariant (diffeomorphism invariant) theory.  Following  the Rauch comparison theorem it is easy to 
argue that a manifold with more negative sectional curvature will have geodesics diverging faster from each other than one with less negative, hence larger, 
sectional curvature.  This is a different way of thinking about the geometric meaning of sectional curvature as compared to (3). The more difficult point 
is to prove that the variations in curvature varying continuously from point to point and in different 
directions leave the rates of expansion and contraction of the appropriate stable and unstable distributions essentially unchanged. To this end, one can
use the rigidity of the hyperbolic space (manifold of constant negative curvature $K<0$) and the properties of its ideal boundary [20],[29].\\

Now we can see why the largest  Lyapunov exponent of a system described by 
the Tsallis entropy is zero. As was pointed out in previous work [30], [31] an effective framework in which the Tsallis entropy can be understood, and 
where 
 nice algebraic properties with respect to the Tsallis entropy composition arise, is provided by constructing a deformation of the reals by the entropic 
 parameter \ $q$ \  which results in the set \ $\mathbb{R}_q$. \ The field isomorphism \ $\tau_q : \mathbb{R} \rightarrow \mathbb{R}_q$ \ is given by [30],
 [31] 
 \begin{equation}
  \tau_q(x) = \frac{(2-q)^x - 1}{1-q}
 \end{equation}
 and we provided some preliminary comments on the induced metric properties of such \  $\tau_q$ \ for \ $q\in (0,1)$. \ 
 We saw in [30] (see also [31]) that the importance of such a map may not so much its exact form, as much as that that it is of exponential form. 
 What \ $\tau_q$ \ does is that it maps distances on \ $\mathbb{R}^n$ \ to exponentials of such distances in $\mathbb{R}_q^n$. \  From the 
 viewpoint of the Cartesian product \ $\mathbb{R}_q^n$ \ all distances of \ $\mathbb{R}^n$ \ appear to have shrunk, as they appear to be measured in a 
 logarithmic scale. Equivalently, what \ $\tau_q$ \ does is that it uses exponentially large scales to measure distances on any Euclidean space \ 
 $\mathbb{R}^n$, \ and by extension on any Riemannian manifold. 
 Such an exponential scale which is used to measure distances is natural, from the viewpoint of the composition properties of theTsallis entropy. Then 
 any exponentially increasing quantity, such as the separation of nearby geodesics on manifolds with negative sectional curvature, will appear to have a 
 polynomial (power-law) asymptotic behavior.  As a result, from the viewpoint of the metric structure induced by the Tsallis entropy, all geodesics on 
 Riemannian  manifolds of negative sectional curvature separate with distances which increase at polynomial rates. Hence the resulting largest Lyapunov 
 exponent of the geodesic flow is zero, a behavior coined ``weak chaos" [2] (and references therein). This is the kind of dynamical behavior that the Tsallis 
 entropy is conjectured to encode. For preliminary indications that this may indeed be so, see [2], [11] and references therein. To our knowledge, 
 no exceptions are currently known to the fact that the Tsallis entropy describes systems with zero largest Lyapunov exponent.  Following the picture 
 presented above, if someone wished to have positive largest Lyapunov exponent with respect to the metric induced by \ $\tau_q$ \ from the Tsallis 
 entropy, then nearby geodesics would have to separate at rates which are increasing at least as fast as \ $ e^{e^{\chi_x(t)}} $. \ This cannot occur 
 for manifolds of constant negative sectional curvature, as is immediately obvious from (8). \\

                                                                        \vspace{5mm}


   \centerline{\large\sc 4. \ Generalized geometric structures and discussion}

                                                                     \vspace{3mm}
 
 The preceding section justifies why the largest Lyapunov exponent is zero, for systems described by the Tsallis entropy.   
 Hence if one persists in quantifying the asymptotic behavior of geodesics, then a modification of the definition of Lyapunov exponents is required.
 Since the exponential growth of nearby perturbations is a very strong condition as seen above, a polynomial (power-law) rate of increment may be 
 more appropriate. To capture this rate of growth, modified Lyapunov exponents can be defined by      
 \begin{equation}
        \log \lambda_x(Y) = \lim_{t\rightarrow\infty} \frac{1}{t} \log ||d_xf_tY||
 \end{equation}
Such modified exponents encode perturbations growing polynomially as \ $t^{\lambda_x(Y)}$. \  This is in agreement with the conjectured 
behavior [2], [7]-[10]
\begin{equation} 
   \frac{d\xi}{dt} = \lambda_q \xi^q  
\end{equation}
for systems described by the Tsallis entropy.
Naturally, these modified Lyapunov exponents will explicitly depend on the value of the non-entropic parameter  $q$.  Since, as is conjectured, the 
systems described by the Tsallis entropy may possess more than just one value of the non-entropic parameter $q$ \ [2],[11], 
depending on which property of  the system is described, such modified Lyapunov exponents would explicitly depend on the value of 
$q$ characterising the sensitivity  of the system to infinitesimal perturbations \ $q_{\mathrm{sen}}$ \ [2].  \\

For this kind of definition to be of physical interest though, there must exist  physical systems exhibiting such a behavior. 
 To be more precise,  we would like to know whether there are geometric structures which are the configuration or phase spaces of systems the 
 stability of the geodesic flow of which is characterized by the modified Lyapunov exponents.  Such structures cannot be Riemannian manifolds, as was 
 established in the previous section, since what we are seeking are systems whose perturbations grow according to a polynomial (power) law. 
Additionally, we would like to understand how do such structures arise in cases of physical interest. \\   

 First, a generalization of the concept of minimal geodesic:  let  \ $(M, d)$ \ be a metric space and let \ $x, y \in M$. \ Then a geodesic between \ 
 $x$ \ and \ $y$ \  is an isometric embedding \ $c$ \ of the interval \ $[0, d(x,y)]$ \ into \ $M$ \ such that \ $c(0) = x$ \ and \ $c(d(x,y)) = y$. \  
 To proceed, one makes  simple but 
 fundamental observation:  when  Riemannian manifolds have positive sectional curvature then triangles whose sides are locally minimal geodesics 
 are``fatter" than corresponding triangles in Euclidean space. The triangles in both spaces are assumed to have corresponding sides of the same 
 lengths. 
 When such manifolds have negative sectional curvature then the corresponding triangles are ``thinner" than in the Euclidean case. The comparison can 
 be 
 equally well be made with a space form of sectional curvature \ $k\neq 0$ \ instead with the Euclidean space [17] - [20]. When, for instance, a manifold 
 has 
 sectional curvature smaller than $k<0$ (which is the case of interest in what follows), then its geodesic triangles will be ``thinner" than those of the space 
 form of sectional curvature  $k$, when both triangles have the same corresponding side-lengths.  A.D. Alexandrov [32], [33] reverses this argument by 
 postulating the ``fatter" or ``thinner" triangles condition in order to define curvature in a comparison sense. What one gains by adopting such a definition 
 is  
 that it applies without a pre-requisite of smoothness of the underlying structure, nor of even a local Euclidean behavior such that of a topological 
 manifold.  
 A metric tree, namely a graph without loops, where each edge  connecting two vertices is assigned a unit length is a standard example of such a 
 structure, 
 which does not look like a manifold at all, but for which curvature in the comparison  sense can still be defined [34], [20].  
 What we lose by using such a definition is that we cannot speak about a space whose 
 sectional curvature is of a particular value but only about a space whose sectional curvature is smaller (or larger) than a particular, negative (for the case 
 of our interest), value.\\
  
 By ``spaces" we will mean metric spaces with intrinsic length structures, also called geodesic spaces, although more general constructions are possible 
 [20]. 
 The geodesic space requirement is that for any two points of such a space there is at least 
 one minimal geodesic joining them. For the case of Riemannian manifolds of negative sectional curvature such a geodesic is actually unique [17]-[20],
 [24],
 [29].  Spaces with an upper bound  \ $k\in\mathbb{R}$ \ on their curvature, which is defined in a comparison sense,  are called \ $CAT(k)$ \  spaces [34],
 [20]. 
 In the case of metric trees, mentioned above, one can easily check  that they are indeed \ $CAT(0)$ \ spaces [34]. 
 Other examples of \ $CAT(k)$ \  spaces, are of course, 
 Riemannian manifolds of  sectional curvature bounded above by \ $k$. \  There is still some ``rigidity"  in \  $CAT(k)$ \ spaces, else not many general 
 statements could be made about  them. Consider, for instance, a linear space endowed with a norm, which is also \ $CAT(k)$ \ for \ $k\in\mathbb{R}$. \ 
 Then 
 this turns out to be an inner product space.  On the other hand, some properties of \ $CAT(k)$ \ spaces may be unexpected. The local to global principle 
 does not apply for such spaces, for instance:  an infinitesimal curvature bound does not necessarily imply a global curvature bound.  
 For a \ $CAT(k)$, \ $k<0$ space, using the triangle inequality, one can prove that the asymptotic separation of geodesics is either bounded or 
 exponential [33], [34]. Nothing new or unexpected  occurs here  when compared to the Riemannian case. But, in stark contrast to the Riemannian case, it 
 is possible to  have a quadratic deviation of geodesics in a \  $CAT(0)$ \ space.  The exact statements and the associated construction can be found in 
 [35]\footnote{\noindent We would like to thank Professor Panos Papasoglou for bringing this reference to our attention} 
 In such cases the modified Lyapunov exponents defined in  the previous section may be of interest in  capturing the asymptotic 
 behavior of the geodesic flow.  In [36] we presented in more detail some of the above constructions and argued in greater detail, 
 why \ $CAT(k)$ \ spaces are indeed useful for describing interacting systems having different values of the entropic parameter.  \\
 
 Why would such generalized metric structures be of interest to Statistical Mechanics? Shouldn't Riemannian manifolds and their asymptotic geodesic 
 behavior be sufficient? The answer is two-fold: First, it is indeed true that configuration and phase spaces are very well modelled by Riemannian
 manifolds and their (iterated) tangent and cotangent spaces as well as  their tensor products. A chance for something more general to be needed comes 
 from at  least two sources: one is when the symmetries of the system impose constraints which introduce singularities to these Riemannian manifolds. A 
 typical case is the non-free action of a group on the configuration/phase space. This is not an unusual occurrence in multi-particle systems and is even 
 more 
 pervasive in the statistical description of classical and quantum fields. Another case is the case of lattice models such as the Ising, Potts etc. There, the 
 configuration space is discrete  so it requires an approach somewhat different from the one applied for continuous systems. Another potential reason 
 for  use of genelarized geometric structures is the thermodynamic limit. The exact way of even appropriately defining the thermodynamic limit tends to be 
 quite non-trivial, especially for systems with long-range interactions [2],[27]. It is not obvious that the thermodynamic limit of such a system will still be a 
 Riemannian space. We can illustrate this with a simple, pictorial, example of a limiting process. Consider a two dimensional cone whose apex has been 
 rounded.  This is a Riemannian manifold. Consider now a sequence in which a neighborhood of the apex gets more and more sharp. 
 The limit of such a sequence will be the cone, which however is not a Riemannian manifold at its apex. Notice that we have avoided 
 providing any details regarding the nature and definition of such a limit, by just alluding to a picture. Such an argument can be made though for a 
 variety of limiting procedures, although the one that we have in mind in this case is the pointed Gromov-Hausdorff convergence [20],[37].   
 By analogy, we may be forced to consider generalized metric structures as the  (thermodynamic) limits of Riemannian configuration/phase spaces. 
 The second source of non-Riemannian geometric structures is an obvious one: 
 the majority of dynamical systems for which a statistical description may be of interest are not Hamiltonian or even dissipative modifications of 
 Hamiltonian systems. For these, the Tsallis entropy seems to be quite useful in encoding the asymptotic (long-time) features of their orbit structure [2]  
 even though such systems may not be described by smooth structures on Riemannian manifolds. Someone can readily mention the fractal-like 
 behavior in the phase space of  systems like these, which also motivated the introduction of the Tsallis entropy in the first place
 [1],[2].  After all, many (if not most) configuration spaces of systems of 
 physical  interest are actually spaces of mappings where it is the highly irregular mappings that provide the emerging behavior. 
 An example of this case are the normal (or  generalized)  random walks, and by direct extension, the field theoretical treatment of quantization/
 thermalization of systems via the  functional integral formalism.\\    
 
 A question directly related to the present work, is why the Tsallis entropy induces such an exponential increase in distances through \ $\tau_q$ \ 
 which,  in turn, results in the largest Lyapunov exponent of the systems described by it to be zero. What is the root cause of such a behavior? 
 Where exactly is this type of hyperbolic behavior, where distances are measured in an exponential scale, is encoded in the Tsallis entropy? Is this 
 due to the particular metric properties of the field isomorphism \ $\tau_q$ \ that was constructed in [30],[31] or is this a generic behavior of the Tsallis 
 entropy which just becomes manifest by such a construction of \ $\tau_q$? The answer is the latter: the exponential expansion of the scale in which 
 the distances are measured is already built in the Tsallis entropy. The construction of \ $\tau_q$ \  [30],[31] is just one particular way to exhibit this 
 and its associated properties. That  this kind of hyperbolic behavior can be expected, is already encoded in the composition/generalized additivity 
 property of the Tsallis entropy itself. How exactly this happens, a fact which is also loosely related to aspects of the uniqueness of the Tsallis \
 entropy, will be discussed extensively in a near-future work.\\       
 
 There is also a more general issue that may be of concern for some systems, such as systems with long-range interactions. In many such cases 
 a cross-over behavior is observed in simulations of dynamical systems with few degrees of freedom [38] - [40]: such systems appear to be have 
 power-law sensitivity to the initial conditions for a relatively long time, a situation dubbed as ``quasi-stationary". Such a behaviour can be described 
 by the Tsallis entropy as ia well-known. After a long time passes though, such systems exhibit an exponential sensitivity the initial conditions, a behaviour
 that can be described by the BGS entropy. In essence one observes that such systems 
 transit from weakly chaotic to strongly chaotic regions, depending on the elapsed time. Such cross-over in the sensitivity behaviour of the initial 
 conditions begs for an explanation. The Lyapunov exponents are  ill-adapted to explaining such a complex behaviour because they are asymptotic invariants.
  Hence their calculation should ignore any transient behaviour that may be present during the evolution of the dynamical system, no matter how long 
 this finite transient behaviour may last. To deal with such a complex behaviour one may have to accept that the nonextesive  sensitivity \ $q_{sen}$ \ may be a 
 function of the evolution parameter (``time"). In the simplest case, and in order to retain some predictive power of this formalism, \ $q_{sen}$ \ should be chosen 
 to be piecewise constant in each of the different temporal regimes of interest. It then becomes part of the theory to not only predict such values for \ $q_{sen}$ \ 
 but also to determine the corresponding time intervals for which such values are effectively constant. It is unclear to us, at the present time, how much one 
 would be able to proceed along these lines and to what extent this approach would give results consistent with the numerical data, so it may be a direction 
 of investigation worth pursuing.\\
  
\noindent  {\sc Acknowledgement:} \ The author is grateful to Professor C. Tsallis for emphasizing that it is the largest Lyapunov exponent that 
vanishes in  systems described by the Tsallis entropy, but not all of them, and for bringing to his attention reference [31] 
of which the author was unaware during the development and writing of the present manuscript as well as that of [30]. The author wishes 
to express his gratitude to the referee for his constructive criticism of the manuscript  and for bringing to the author's attention references [38] - [40].  \\

                                     \vspace{8mm}


      \centerline{\large\sc References}
      
                                \vspace{5mm}

\noindent [1] \  C. Tsallis, \  J. Stat. Phys. {\bf 52}, \ 479 \  (1988)  \\
\noindent [2] \  C. Tsallis, \ Introduction to Nonextensive Statistical Mechanics: \\
                           \hspace*{6mm}  Approaching a Complex World, \ Springer (2009)\\
\noindent [3] \ L. Boltzmann, \  Acad. Wissen. Wien, Math.-Naturwissen. {\bf 75}, \  67 \ (1877)\\
\noindent [4] \ A. Einstein, \ Ann. der Phys., {\bf 33}, \ 1275 \ (1910)\\
\noindent [5] \ E.G.D. Cohen, \ Pramana {\bf 64}, \ 635 \ (2005)\\
\noindent [6] \ G. Gallavotti, \ Statistical Mechanics: A Short Treatise, \ Springer \ (1999)\\
\noindent [7] \ P. Grassberger, M. Scheunert, \ J. Stat. Phys. {\bf 26}, \ 697 \ (1981)\\
\noindent [8] \ G. Anania, A. Politi, \ Europhys. Lett. {\bf 7}, \ 119 \ (1988)\\
\noindent [9] \ H. Hata, T. Horita, H. Mori,  \ Prog. Theor. Phys. {\bf 82}, \ 897 \ (1989)\\
\noindent [10] \ H. Mori, H. Hata, T. Horita, T. Kobayashi, \  Prog. Theor. Phys. Suppl. {\bf 99}, \ 1 \ (1989)\\
\noindent [11] \ C. Tsallis, \ Some Open Points In Nonextensive Statistical Mechanics, \ arXiv:1102.2408v1\\
\noindent [12] \ C. Tsallis, A.R. Plastino, W.-M. Zheng, \ Chaos, Solitons and Fractals {\bf 8}, \ 885 \ (1997)\\   
\noindent [13] \  F. Baldovin, A. Robledo, \ Europhys. Lett. {\bf 60}, \ 518 \ (2002)\\ 
\noindent [14] \  F. Baldovin, A. Robledo, \ Phys. Rev. E {\bf 66}, \ R045104 \ (2002)\\
\noindent [15] \  E. Mayoral, A. Robledo, \ Phys. Rev. E {\bf 72}, \ 026209 \ (2005)\\
\noindent [16] \  A. Robledo, \ Physica A {\bf 370}, \ 449 \ (2006)\\
\noindent [17] \ J. Cheeger, D.G. Ebin, \ Comparison Theorems in Riemannian Geometry, \\
                     \hspace*{8mm} AMS Chelsea \  (1975)\\ 
\noindent [18] \ T. Sakai, \ Riemannian Geometry, \ Amer. Math. Soc. \ (1996)\\
\noindent [19] \ M. Gromov, \ Milan J. Math. {\bf 61}, \ 9 \ (1994)\\
\noindent [20] \ M. Gromov, \ Metric Structures for Riemannian and Non-Riemannian Spaces,\\
                     \hspace*{8mm}  Birkh\"{a}user \ (1999)\\                         
\noindent [21] \ J.W. Gibbs, \ Elementary Principles in Statistical Mechanics,  Yale Univ. Press (1948) \\              
\noindent [22] \ E.T. Jaynes, \ Foundations of Probability Theory and Statistical Mechanics, \ in \\
                          \hspace*{8mm}  Delaware Seminar in the Foundations of Physics, \ M. Bunge (Ed.), \ Springer (1967)\\ 
\noindent [23] \ E.T. Jaynes, \ Gibbs vs. Boltzmann Entropies, \  Amer. J. Phys. {\bf 33},  \  391 \ (1965)\\ 
\noindent [24] \ A. Katok, B. Hasselblatt, \ Introduction to the Modern Theory of Dynamical \\ 
                    \hspace*{8mm} Systems, \  Cambridge Univ. Press \ (1995)\\                   
\noindent [25] \  L. Barreira, Y. Pesin, \ Dynamics of Systems with Nonzero Lyapunov Exponents, \\  
                    \hspace*{8mm} Cambridge Univ. Press \ (2007)\\    
\noindent [26] \ A. Lederhendler, D. Mukamel, \ Phys. Rev. Lett. {\bf 105},  \ 150602 \ (2010)\\ 
\noindent [27] \ F. Bouchet, S. Gupta, D. Mukamel, \  Physica A {\bf 389}, \ 4389 \ (2010)\\
\noindent [28] \ M. Costeniuc, R.S. Ellis, H. Touchette, B. Turkington, \ Phys. Rev. E {\bf 73}, \ 026105\\
                     \hspace*{8mm} (2006)\\
\noindent [29] \ G. P. Paternain, \ Geodesic Flows, \ Birkh\"{a}user \ (1999)\\    
\noindent [30] \ N. Kalogeropoulos, \  Physica A {\bf 391}, \ 1120 \ (2012)\\
\noindent [31] \ T.C. Petit Lob\~{a}o, P.G.S. Cardoso, S.T.R. Pinho, E.P. Borges, \  Braz. J. Phys. {\bf 39},\\
                          \hspace*{8mm}   402 \  (2009) \\                        
\noindent [32] \ A.D. Alexandrov, \ Trudy Mat. Inst. Steklov {\bf 38}, \ (1951)\\ 
\noindent [33] \ V.N. Berestovskij, I.G. Nikolaev, \ Multidimensional Generalized Riemannian Spaces,\\  
                    \hspace*{8mm}  in  \emph{Geometry IV: Nonregular Riemannian Geometry}, Y.G. Reshetnyak (Ed.), Encycl.\\
                    \hspace*{8mm}  Math. Sci., \  Vol. {\bf 70},  \ Springer \ (1993)\\
\noindent [34] \ M. Gromov, \ Asymptotic Invariants of Infinite Groups, in \emph{Geometric Group Theory, \\
                   \hspace*{8mm}  Vol. 2,}   G. Niblo, M. Roller (Eds.), \ Lond. Math. Soc. Lecture Notes Series \ {\bf 182},\\
                   \hspace*{8mm}  Cambridge Univ. Press \ (1993)\\ 
\noindent [35] \  S.M. Gersten, \ Geom. Funct. Anal. {\bf 4}, \  37 \ (1994)\\   
\noindent [36] \ N. Kalogeropoulos, \ Physica A {\bf 391}, \ 3435 (2012)\\
\noindent [37] \ K. Fukaya, \  Adv. Studies Pure Math. {\bf 18}, \ 143 \ (1990)\\
\noindent [38] \ F. Baldovin, C. Tsallis, B. Schulze, \ Physica A {\bf 320}, \ 184 \ (2003)\\
\noindent [39] \ F. Baldovin, E. Brigatti, C. Tsallis, \ Phys. Lett. A {\bf 320}, \ 254 \ (2004)\\
\noindent [40] \ G.F.J. A\~{n}a\~{n}os, F. Baldovin, C. Tsallis, \ Eur. Phys. Jour. B {\bf 46}, \ 409 \ (2005)\\
\end{document}